\documentclass[twocolumn,aps,prb,superscriptaddress,citeautoscript]{revtex4}
\usepackage{graphicx}
\usepackage{bibunits}

\newcommand{\um}{$\mu$m}
\newcommand{\Flat}{F_{\rm lat}}

\newcommand{\fig}[1]{Fig.~\ref{#1}}

\begin{document}
\begin{bibunit}
\author{E. W. J. Straver}
\affiliation{Departments of Applied Physics and Physics, and Geballe Laboratory for Advanced Materials, Stanford University, Stanford, CA 94305}
\author{J. E. Hoffman}
\affiliation{Departments of Applied Physics and Physics, and Geballe Laboratory for Advanced Materials, Stanford University, Stanford, CA 94305}
\affiliation{Department of Physics, Harvard University, Cambridge, MA 02138}
\author{O. M. Auslaender}
\email[Corresponding author: ]{ophir@stanford.edu}
\affiliation{Departments of Applied Physics and Physics, and Geballe Laboratory for Advanced Materials, Stanford University, Stanford, CA 94305}
\author{D. Rugar}
\affiliation{IBM Research Division, Almaden Research Center, 650 Harry Road, San Jose, CA 95120}
\author{Kathryn A. Moler}
\affiliation{Departments of Applied Physics and Physics, and Geballe Laboratory for Advanced Materials, Stanford University, Stanford, CA 94305}

\title{Controlled Manipulation of Individual Vortices in a Superconductor}

\begin{abstract} 
We report controlled local manipulation of single vortices by low temperature magnetic force microscope (MFM) in a thin film of superconducting Nb. We are able to position the vortices in arbitrary configurations and to measure the distribution of local depinning forces. This technique opens up new possibilities for the characterization and use of vortices in superconductors.
\end{abstract}

\maketitle
Quantized magnetic flux tubes, called vortices, allow superconductivity to survive in high applied magnetic field in type-II superconductors, making them more technologically relevant than type-I. Each vortex has a non-superconducting core with a radius on the scale of the coherence length $\xi$, and a circulating supercurrent that generates one quantum of magnetic flux, $\Phi_0=h/2e$, over the scale of the London penetration depth $\lambda$. Although the paired electrons in a superconductor carry charge without resistance, a current will exert a magnus force on all vortices, which results in dissipation, if any of them move. Vortex motion is both a challenge and an opportunity. The challenge is to understand and reduce uncontrolled vortex motion. A vortex may be pinned in place by co-locating its energetically costly non-superconducting core with a defect that locally suppresses superconductivity. Decades of materials research have characterized pinning strengths and engineered defects to increase pinning.\cite{Larbalestier01,Scanlan04,Haugan04,Kang06} Continued reduction in uncontrolled vortex motion will open up new applications, both for quiet circuits in sensing and communication, and for large currents in high-field magnets and power distribution.

Controlled vortex motion, on the other hand, has great prospects for logic applications and for fundamental science. Collectively controlled vortex motion can serve as a rectifier,\cite{Villegas03} a vortex ratchet mechanism can perform clocked logic,\cite{Hastings03} and vortices can control spins in an adjacent diluted magnetic semiconductor,\cite{Berciu05} while vortices adjacent to an electron gas in a quantum-Hall state may allow the creation of exotic quantum states.\cite{Weeks07} A proposed test\cite{OlsonReichhardt04,Nelson04} of the long-standing idea that vortices may entangle like polymers\cite{Nelson88} requires controlled local manipulation of single vortices. Vortices are of theoretical interest for their own sake,\cite{Blatter94,Nelson02} as clues to the underlying superconductivity,\cite{Wynn01,Hoffman02} as analogs for interacting bosons \cite{Kafri07} and as model systems for soft condensed matter.\cite{Nelson02}

Previous experimental manipulations of single vortices have applied relatively delocalized forces\cite{Allen89,Park92,Stoddart95,Breitwisch00,Gardner02} or have not controlled the vortex motion.\cite{Moser98,Roseman02a} Here we demonstrate vortex manipulation with nanoscale control and show that we can quantitatively measure the local depinning force. In magnetic force microscopy (MFM), the sample exerts a measurable force on a cantilever with a sharp magnetic tip, such that scanning the cantilever at a constant height $z$ above the sample provides a map of magnetic features. MFM has been used for a variety of vortex experiments.\cite{Moser98,Roseman02a,Lu02,Volodin02,Pi04b,Ophir08} Many experiments image the cantilever deflection, which is proportional to the vertical force, $F_z$. For improved signal-to-noise ratio, we use frequency modulation (FM) mode,\cite{Albrecht91} in which the imaging signal is a shift $\Delta f$ in the cantilever's resonant frequency, $f_0$. The images show the variations in the derivative of $F_z$, $\partial F_z/\partial z = -2k\Delta f/f_0$, where $k$ is the cantilever's spring constant.\cite{SuM} We use the lateral component of the force $\Flat\equiv|F_{x}{\hat x}+F_{y}{\hat y}|$ to pull or push vortices.
\begin{figure}[tb]
\includegraphics[angle=0,width=3.375in]{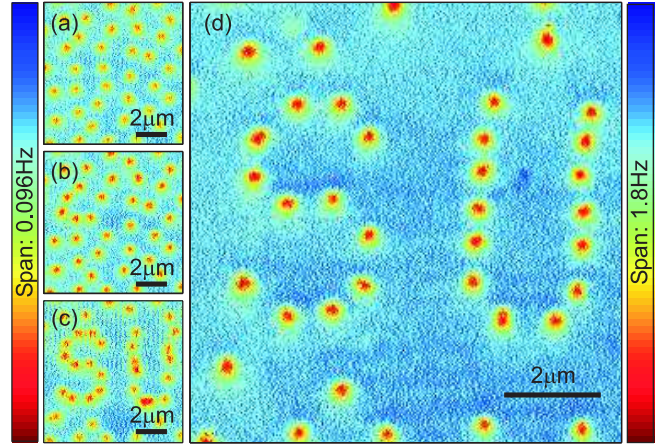}
\caption{\label{fig_su}Manipulation of vortices to spell "SU." Colorbars give $\Delta f$ [left for (a-c), right for (d)]. {\bf(a)} An unmanipulated configuration of pinned vortices after initial cooling to $T=7.0$K, imaged at a scan height $z = 300$nm. {\bf(b-c)} Intermediate configurations after manipulation of some vortices in the temperature range $7.0$ to $7.2$K, imaged at $z = 300$nm. {\bf(d)} Final configuration after completing the vortex manipulation at $7.2$K, imaged at $z = 120$nm and $T = 5.5$K for better resolution and stronger pinning.}
\end{figure}

We used a home-built variable-temperature MFM\cite{Straver04} to study a 300nm thick Nb film sputtered onto a silicon substrate.\cite{HYPRES} The midpoint transition temperature is $T_c = 8.6$K with a transition width $\Delta T_c = 0.6$K, as measured by magnetic susceptibility, and $\lambda = 90$nm.\cite{Radparvar}
\begin{figure}[b]
\includegraphics[width=3.375in]{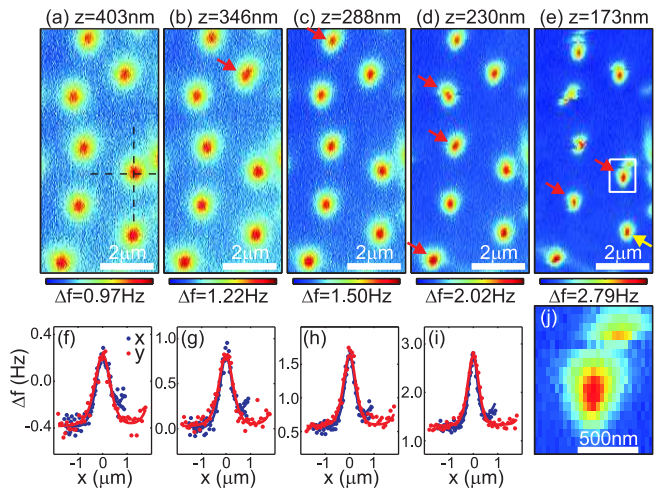}
\caption{\label{fig_depin}Vortices in a Nb film imaged at $T = 5.5$K as a function of decreasing scan height $z$. {\bf(a-e)} Images at 5 different values of $z$. Red right arrows indicate vortices depinning for the first time. A single vortex remains unmoved even at the largest applied force in (e) (yellow left arrow). {\bf(f-i)} Cross-sections along the crosshairs in (a) for the first four $z$-values. The lines are from the monopole-monopole model. {\bf(j)} Closeup on the boxed region in (e).}
\end{figure}

\begin{figure*}[bt]
\includegraphics[angle=0,width=6.750in]{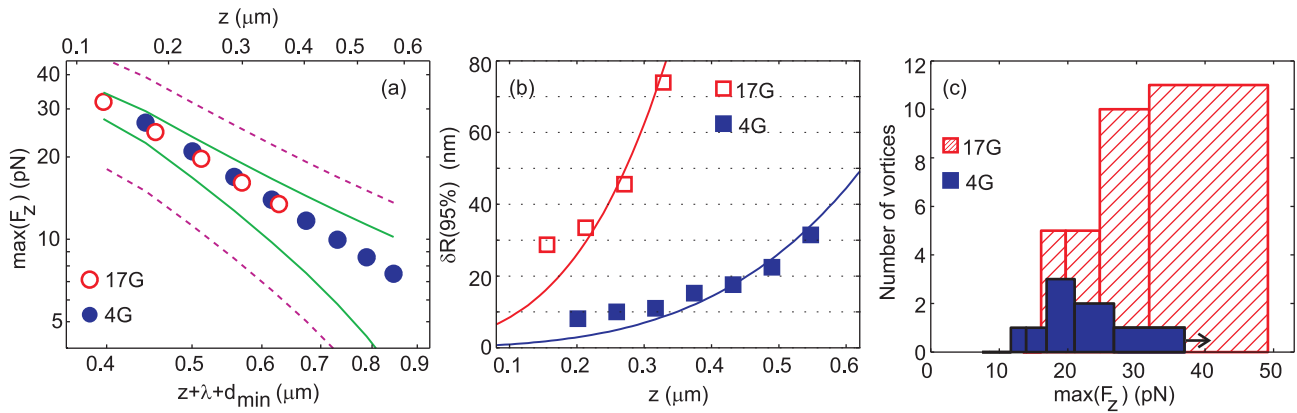}\vspace{-0.25cm}
\caption{\label{fig_force}{\bf(a)} Maximum of the vertical force applied to a vortex, $\max(F_z)$, as a function of the measured tip-sample separation, $z$, and the modeled monopole-monopole separation, $z+\lambda+d$, for two data sets taken with the same cantilever. Solid green lines indicate the systematic error associated with modeling the data. Dashed purple lines also include the uncertainty from the cantilever spring constant. {\bf(b)} The threshold for detecting vortex motion at the 95\% confidence level. The $4$G dataset has better resolution because the data was less noisy. The curves are from the monopole-monopole model.\cite{SuM} {{\bf(c)}} Histogram of number of vortices depinned vs. $\max(F_z)$. The width of the histogram bars is the difference in force between heights of successive scans. The arrow on the last blue bin indicates that a single vortex did not move at the largest applied force, as shown in \fig{fig_depin}.}
\end{figure*}
Figure~\ref{fig_su} shows controlled vortex manipulation. We cooled the sample to $T = 7$K in an external field of a few Gauss with a polarity giving an attractive tip-vortex force. Figure~\ref{fig_su}a shows a disordered arrangement of vortices pinned in the sample. The lack of observed vortex motion during the scan indicates that at this height of $z=300$nm we are imaging vortices without depinning them. To move a vortex, we position the cantilever tip over it, reduce the oscillation amplitude, descend to $z \approx 10$nm, move parallel to the surface, and then withdraw vertically, leaving the vortex at a new location. Some locations require several attempts at different heights or lateral offsets to move the vortex, or different pulling directions and slightly elevated temperature to reduce the pinning force. The exact final location of the vortex must depend on the local pinning potential; pinning sites appear to be dense in the Nb film. Images (\fig{fig_su}b-d), each taken after several manipulations, show the successful repositioning of vortices to write "SU." This procedure is akin to atom manipulation in scanning tunneling microscopy.\cite{Eigler90}

We also quantified the forces required to depin vortices at $5.5$K. Figure~\ref{fig_depin} shows a
series of images, with decreasing $z$, of 8 isolated vortices at a vortex density corresponding to $4$G with a polarity giving repulsive tip-vortex force. As $z$ decreases, the increased $\partial F_z/\partial z$ leads to increased signal strength, while the increasing $\Flat$ depins vortices. Vortices at different locations depin at different heights, indicating a distribution of depinning forces. In this case, $\Flat$ is weak enough that a depinned vortex simply finds a better pinning site nearby, but we have also applied larger forces to sweep the field of view clean of vortices.

We fit the data in \fig{fig_depin}, as well as a second dataset of $31$ vortices at a vortex density corresponding to $17$G, also at $5.5$K. We model the cantilever tip and each vortex as monopoles.\cite{SuM} Cuts through the fits are shown in \fig{fig_depin}(f-i). The resulting $\max(F_z)$ (\fig{fig_force}a) has systematic error bars of less than 50\% from the modeling and from the uncertainty in $k$. These systematic errors do not affect the relative distribution of depinning forces.

Most vortices clearly moved by visual inspection during one or more scans. To quantify our detection threshold, we bootstrap\cite{SuM,Efron93} the entire $8$-vortex data set and one quadrant of the $31$-vortex data set.\cite{SuM} The upper limit to undetected vortex motion at the 95\% confidence level, $\delta R$, is shown in \fig{fig_force}b. The stationary vortex in \fig{fig_depin} did not move by more than $9$nm.

Identifying the $z$-value at which each vortex moved gives a histogram of vertical depinning forces (\fig{fig_force}c) ranging from $12$pN to more than $32$pN. Since it is actually $\Flat$ that causes the depinning, we use modeling to estimate it based on the measured $\max(F_z)$. Depending on the model, the ratio between the maximum $\Flat$ and the maximum $F_z$ ranges from $0.3$ for a pyramidal tip less sharp than ours\cite{Wadas92} to $2/ 3\sqrt3\approx0.38$ in the monopole–monopole model. Using $0.38$, the observed lateral depinning force ranges from $4$pN to above $12$pN, or $15$pN/\um\ to $40$pN/\um\ normalized by film thickness. This technique could be advanced further by using vertical cantilevers with lower spring constants to directly measure the lateral force.

The manufacturer-specified critical current for similar films is $50\pm30$mA per transverse \um\ at 4.2K,\cite{HYPRES} equivalent to a depinning force of $104 \pm 69$pN. Other single vortex pinning measurements in Nb used a transport current to supply a force\cite{Allen89,Park92,Stoddart95,Breitwisch00} and various stationary probes to detect motion. For comparison, we normalize previous results by the film thickness and extrapolate to our reduced temperature using the power law $F \propto (1-T/T_c)^\gamma$, where $\gamma$ is an experimentally determined exponent that varies greatly between experiments. The inferred depinning forces are $123$pN/\um,\cite{Breitwisch00} $80$pN/\um,\cite{Allen89} $58$pN/\um,\cite{Park92} $5$pN/\um,\cite{Allen89} and $0.6$pN/\um.\cite{Stoddart95} These experiments have motion detection thresholds ranging from a few hundred nanometers to microns, except for [\onlinecite{Stoddart95}], which reports a resolution of $\approx20-40$nm, achieved with an array of stationary Hall probes.

Our best threshold for vortex motion detection is better than $10$nm and is limited by the signal-to-noise ratio. The characteristic scale for changes in the pinning potential is the coherence length $\xi\approx10-20$nm.\cite{Allen89,Volodin02} We can therefore detect all vortex depinning events with a quantitative determination of the locally applied depinning force. Imaging the vortices before, during, and after the depinning has great prospects for correlating pinning with topography, for determining the pinning landscape directly, and for studying single-vortex dynamics.

Acknowledgments: This work was supported by the Packard Foundation and the Air Force Office of
Scientific Research. We thank Lan Luan for discussions.


\end{bibunit}
\onecolumngrid
\pagebreak\pagebreak\pagebreak\pagebreak
\setcounter{equation}{0}
\def\theequation{S\arabic{equation}}
\setcounter{figure}{0}
\def\thefigure{S\arabic{figure}}

\begin{bibunit}
\begin{center}
    \Large{\bf Supplemental Material}
\end{center}

\section{MFM tip for Fig.~\ref{fig_su}}
We acquired the data shown in Fig.~\ref{fig_su} using a MikroMasch silicon cantilever with $f_0 = 81$kHz, $k = 4.0\pm0.6$N/m, and an electron beam deposited (EBD) tip, with an iron film of nominal thickness $40$nm evaporated onto one side of the tip.

\section{MFM tip for Figs.~\ref{fig_depin},~\ref{fig_force},~\ref{fig_sup}}
We acquired the data shown in Fig.~\ref{fig_depin}, and analyzed in Figs.~\ref{fig_force} and \ref{fig_sup}, using a Veeco Instruments silicon cantilever (model MESP) with $f_0 = 71$kHz, $k = 2.1\pm0.7$N/m, and a nominally uniform CoCr Veeco-proprietary coating.

\section{Analysis}
\begin{figure}[b]
\includegraphics[angle=0,width=6.75in]{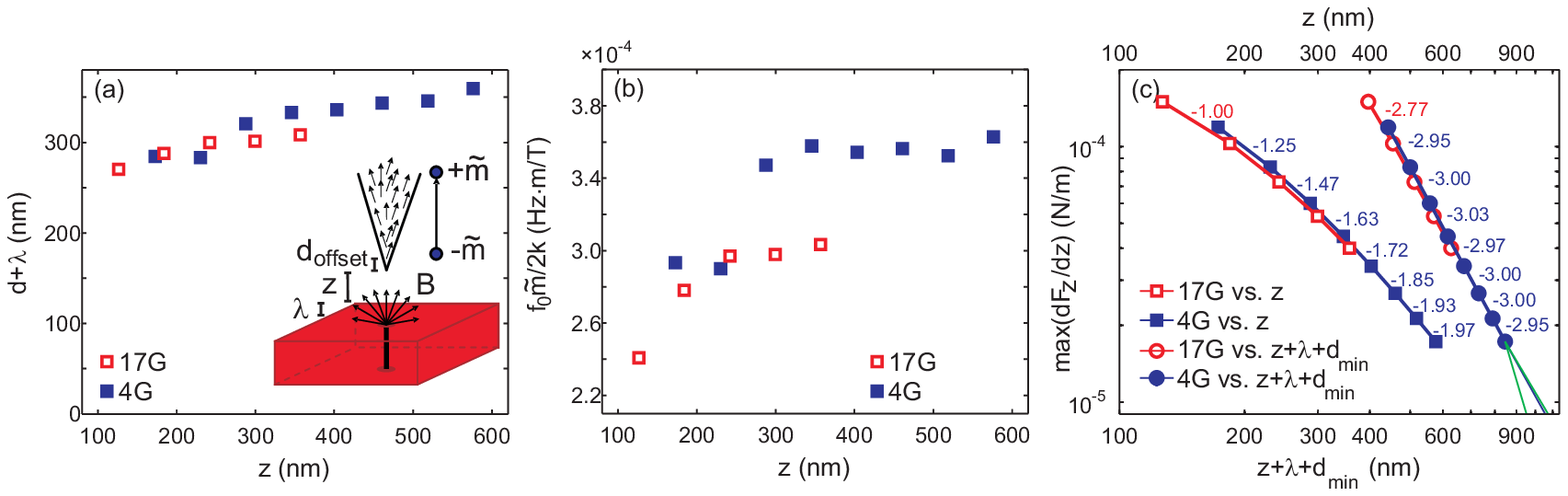}
\caption{\label{fig_sup}{\bf(a)} Fit parameter $d+\lambda$ from the monopole-monopole model, averaged over all as yet unmoved vortices at a given height, vs. $z$. Inset: sketch motivating the monopole model for the tip. {\bf(b)} Fit parameter $f_0{\tilde m}/2k$, also averaged over all as yet unmoved vortices at a given height, vs. $z$. {\bf(c)} The maximum vertical force gradient $\max(\partial F_z/\partial z)$ at each height, plotted against the measured $z$ and against $z+\lambda+d$, with $\lambda+d$ obtained from the fit at the lowest $z$-value. The latter plot shows the expected power law behavior with $p = 3$; the $p$-value for each pair of points is indicated.}
\end{figure}

The field lines for an isolated vortex in a film of finite thickness $t$ and infinite width are given in [\onlinecite{Chang92}]. For $z,t \gg \lambda$, the vortex field lines are those of a monopole of strength $2\Phi_0/\mu_0$ located at $\vec R_0\equiv(x_0,y_0)$ a distance $\lambda$ below the surface,\cite{Pearl66} where $\mu_0$ is the magnetic permeability of vaccum. We model the tip as a monopole $\tilde m$ offset from the physical tip of the pyramid by a distance $d$, \cite{Schonenberger90,Wiesendanger92} as shown in the inset to \fig{fig_sup}a. This "monopole –- monopole" model gives a frequency shift:
\begin{equation}\label{mmm}
    \Delta f_m=\frac{f_0}{2k}{\tilde m}\frac{\partial B^{\rm vortex}_z}{\partial z}=\frac{f_0}{2k}\frac{{\tilde m}\Phi_0}{2\pi}\frac{-(\vec R-\vec R_0)^2+2(z+\lambda+d)^2}{\left[(\vec R-\vec R_0)^2+(z+\lambda+d)^2\right]^{5/2}}
\end{equation}
per vortex, where $B^{\rm vortex}_z$ is the vertical component of the magnetic field from a vortex and $\vec R\equiv(x,y)$. We fit an image to a sum of terms $\Delta f_m$, one for each vortex, plus $df_{\rm offset}$, a $z$-dependent, $\vec R$-independent, offset due primarily to the laterally uniform Meissner repulsion of the tip from the sample. The free parameters of the fit are $\vec R_0$ for each vortex and  $df_{\rm offset}$, $f_0{\tilde m}/2k$, $d+\lambda$, all the same for all vortices in the image. The latter two fit parameters (\fig{fig_sup}a,b) varied by about 25\% over the measured $z$ range, perhaps because the signal is increasingly dominated by the very tip of the pyramid for smaller $z$. The fit value of $f_0{\tilde m}/2k$ corresponds to a monopole moment on the tip equivalent to about 5 vortices.

We also tried a dipole model for the tip, and found that the monopole model gave smaller $\chi^2$ and more consistent values for the tip parameters. Possible refinements include modeling the tip as an extended object, or modeling the vortex using the full expressions for a film of finite thickness. However, both the peak frequency shift, which gives the maximum vertical force gradient, and the vortex center, which enables us to identify vortex motion events, are already well determined by the monopole-monopole model. Also, the bootstrapping required to determine the confidence intervals involves hundreds of iterations of simultaneous multi-vortex fits, and would become cumbersome with more complicated models.

The peak frequency shift, averaged over all as yet unmoved vortices at each height, gives the maximum vertical force gradients $\max(\partial F_z / \partial z)$ (\fig{fig_sup}c). If the monopole-monopole model is strictly valid with no $z$ dependence of the parameters, we expect $\max(\partial F_z / \partial z) \propto (z + \lambda + d)^{-p}$ with $p=3$. The value of $p$ for fits to each pair of consecutive $\max(\partial F_z / \partial z)$ points ranges from 2.95 to 3.03 (\fig{fig_sup}c), indicating good agreement. The measured $\max(\partial F_z / \partial z)$ data, extrapolated to infinity with $p = 2.95$ and integrated from infinity to $z$, gives the maximum vertical force $\max(F_z)$ (Fig.~3\ref{fig_force}a).

To characterize the possible systematic error associated with this extrapolation, we note that $\max(\partial F_z / \partial z)$ plotted vs. $z$ on a log-log plot has negative curvature. We
extrapolate using $\max(\partial F_z / \partial z)\propto z^{-1.97}$ for an upper bound, and $\max(\partial F_z /\partial z)\propto z^{-4}$, the power expected for a dipole tip, as a lower bound. This systematic error could be reduced in future experiments by taking additional data at larger $z$. The other main source of systematic error is the uncertainty in the cantilever spring constant.

\subsection{Detection threshold for vortex motion}
The threshold for vortex motion detection is determined by bootstrapping an 8-vortex dataset at $4$G and a 9-vortex dataset at $17$G. In each case, we fit all 8 or 9 vortices simultaneously to Eq.~\ref{mmm}, allowing $f_0{\tilde m}/2k$, $d+\lambda$, $x_0$ and
$y_0$ to vary separately for each vortex. Because $f_0{\tilde m}/2k$ and $d+\lambda$ are properties of the tip, we expect them to be constant for all vortices at a given height; indeed we find that
$f_0{\tilde m}/2k$ varies by less than 2\%, while $d+\lambda$ varies by less than 5\%. We use at least 600 bootstrapping iterations to find the statistical distribution of ($x_0$, $y_0$) pairs for each vortex center at each height. Each vortex has a bootstrapped set of center parameters ${x_0}_{n,v,i}$ and ${y_0}_{n,v,i}$ where $n$ labels the scan height, $v$ labels the particular vortex, and the bootstrap iteration index $i$ runs from 1 to at least 600. To find the confidence levels for the motion detection threshold (Fig.~\ref{fig_force}b), we looked at the distribution of $\delta R=\sqrt{({x_0}_{n,v,i}-{x_0}_{n+1,v,i})^2+({y_0}_{n,v,i}-{y_0}_{n+1,v,i})^2}$.

We have also derived a formula for the detection threshold. For concreteness let us assume that the tip-vortex interaction is indeed described by the monopole-monopole model. The motion detection threshold is proportional to the ratio between noise of the measurement ($\delta\Delta f$) and the slope of the measured curve:
\begin{equation}\label{eq_dRa}
    \delta R=2k\frac{\delta\Delta f/f_0}{\left(\partial/\partial R\right)\left(\partial F_z/\partial z\right)}.
\end{equation}
Assuming that the vortex first moves when the lateral force is maximal, which in the monopole-monopole model occurs for $R^*\equiv(z+\lambda+d)/\sqrt2$, we find:
\begin{equation}\label{eq_dRb}
\delta R=A\left(z+\lambda+d\right)^{4},
\end{equation}
where $A\equiv\frac{3^{7/2}k}{4{\tilde m}\Phi_0}(\delta\Delta f/f_0)$. Equation~\ref{eq_dRb} is the formula plotted in Fig.~\ref{fig_force}b, where we used $\lambda+d$ from a linear fit to \fig{fig_sup}a and proportionality constants, $A$, from fitting the data in Fig.~\ref{fig_force}b to Eq.~\ref{eq_dRb}.


\end{bibunit}

\end{document}